\documentclass[12pt]{article}
\usepackage{amssymb,amsmath}
\usepackage[bf,small,hang]{caption}

\newcommand{\be}{\begin{equation}}
\newcommand{\ee}{\end{equation}}
\newcommand{\La}{\cal{L}}
\newcommand{\no}{\noindent}
\newcommand{\ba}{\begin{eqnarray}}
\newcommand{\ea}{\end{eqnarray}}
\newcommand*{\pd}{\partial}
\newcommand*{\pdm}{\pd_{\mu}}

\newcommand*{\bea}{\begin{eqnarray}}
\newcommand*{\eea}{\end{eqnarray}}
\newcommand*{\pref}[1]{(\ref{#1})}

\newcommand*{\mn}{{\mu\nu}}

\newcommand*{\gh}{\mathrm{gh}}
\newcommand*{\puregh}{\mathrm{puregh}}
\newcommand*{\antifd}{\mathrm{antifd}}
\newcommand*{\sdet}{\mathrm{sdet}}
\newcommand*{\str}{\mathrm{str}}
\newcommand*{\uth}{^\mathrm{th}}

\begin{document}

\title{BRST-antifield Quantization: a Short Review}

\author{\bf{Andrea Fuster$^a$, Marc Henneaux$^{b,c}$ and Axel Maas$^d$}\\
\small $^a$NIKHEF, P.O. Box 41882, 1009 DB Amsterdam, The Netherlands\\
\small $^b$Physique Th\'eorique et Math\'ematique, Universit\'e Libre de Bruxelles, and\\
\small International Solvay Institutes, ULB Campus Plaine C.P. 231,\\
\small  B--1050 Bruxelles, Belgium \\
\small $^c$ Centro de Estudios Cient\'{\i}ficos (CECS), Valdivia,
Chile \\
\small $^d$ GSI, Planckstra{\ss}e 1, 64291 Darmstadt, Germany
\vspace*{1cm}\\}

\date{}

\maketitle

\abstract{Most of the known models describing the fundamental
interactions have a gauge freedom. In the standard path integral, it
is necessary to ``fix the gauge" in order to avoid integrating over
unphysical degrees of freedom. Gauge independence might then become
a tricky issue, especially when the structure of the gauge
symmetries is intricate. In the modern approach to this question, it
is BRST invariance that effectively implements gauge invariance.
This set of lectures briefly reviews some key ideas underlying the
BRST-antifield formalism, which yields a systematic procedure to
path-integrate any type of gauge system, while (usually) manifestly
preserving spacetime covariance. The quantized theory possesses a
global invariance under what is known as BRST transformation, which
is nilpotent of order two. The cohomology of the BRST differential
is the central element that controls the physics. Its relationship
with the observables is sketched and explained. How anomalies appear
in the ``quantum master equation" of the antifield formalism is also
discussed. These notes are based on lectures given by MH at the
10$\uth$ Saalburg Summer School on Modern Theoretical Methods from
the 30$\uth$ of August to the 10$\uth$ of September, 2004 in
Wolfersdorf, Germany and were prepared by AF and AM. The exercises
which were discussed at the school are also included.}

\section{Introduction}
\setcounter{equation}{0}

Gauge symmetries are omnipresent in theoretical physics,
especially in particle physics. Well-known examples of gauge
theories are QED and QCD. A common feature of gauge theories is
the appearance of unphysical degrees of freedom in the Lagrangian.
Because of this, the naive path integral for gauge theories is
meaningless since integrating over gauge directions in the measure
would make it infinite-valued: \be \int D A_{\mu}\; e^{iS}= \infty
\ee \no The redundant gauge variables must be removed from the
theory by considering gauge-fixing conditions. When this is done,
gauge invariance is of course lost and it is not clear how to
control the physics. The modern approach to cope with these
problems in the case of general gauge theories was developed by
Batalin and
Vilkovisky~\cite{Batalin:1981jr,Batalin:1983wj,Batalin:1984jr},
building on earlier work by Zinn-Justin~\cite{zj},
Kallosh~\cite{Kallosh} and de Wit and van Holten~\cite{dWvH} . It
goes under the name of BV or antifield formalism and is the method
explained in these lectures. The BRST symmetry is central to it
\cite{Becchi:1974xu,Becchi:1974md,Becchi:1975nq,Tyutin:1975qk}. A
complete coverage of the topic and further references can be found
in \cite{Henneaux:1989jq,Henneaux:1992ig,Barnich:2000zw}.

\section{Structure of gauge symmetries}
\setcounter{equation}{0} \label{structure}

Let us give a brief overview of the different types of
gauge theories that one may encounter. \\

\subsection{Yang-Mills type}

We are all familiar with Yang-Mills gauge theories. In the absence
of matter, the action is given by \bea
S_0[A_\mu^a]&=&-\frac{1}{4}\int F_{\mu\nu}^aF^{\mu\nu}_ad^nx\\
F_{\mu\nu}^a&=&\pdm A_\nu^a-\pd_\nu A_\mu^a+f^a_{\;\;bc}
A_\mu^bA_\nu^c.\label{nafieldstrength} \eea \noindent
$F_{\mu\nu}^a$ is the field strength tensor, $A_\mu^a$ is the
gauge field and $f^a_{\;\;bc}$ are the structure constants of the
associated gauge group. The dimensionality of space-time is $n$.
The gauge transformation takes infinitesimally the form \bea
A_\mu^a(x)'&=&A_\mu^a(x)+\delta_\epsilon A_\mu^a\\
\delta_\epsilon A_\mu^a(x)&=&D_\mu\epsilon^a(x)=
\pdm\epsilon^a+f^a_{\;\;bc}A_\mu^b\epsilon^c(x), \eea \noindent
where $\epsilon^c(x)$ is a set of arbitrary functions, the gauge
parameters. In these theories, the commutator of infinitesimal
gauge transformations reads \be
[\delta_{\varepsilon},\delta_{\eta}]\;X=\delta_{\xi}X,\;\;
\;\;\xi^a=f^a_{\;\;bc}\varepsilon^b\eta^c \ee \no with
$\varepsilon,\;\eta,\;\xi$ gauge parameters and where $X$ can be
any field. It is clear from this formula that the algebra of the
gauge transformations closes off-shell as the commutator of the
gauge transformations is again a gauge transformation of the same
type, without using the equations of motion.

\subsection{Closure only on-shell}
Off-shell closure holds for Yang-Mills gauge theories but is not a
general feature of gauge systems. The gauge transformations might
close only when the equations of motion hold.  Notable examples
where this is the case are extended supergravity theories.  Rather
than discussing the gauge structure of supergravities, which is
rather intricate, we shall illustrate ``closure only on-shell" in
the case of a much simpler (but of no direct physical interest)
system. Consider the following Lagrangian: \be {\La}
=\frac{1}{2}(\dot{q}^1-\dot{q}^2-\dot{q}^3)^2=\frac{1}{2}
\dot{y}^2,\;\;\;\; y=q^1-q^2-q^3 \label{lagcoff} \ee \no for a
model with three coordinates $q^1$, $q^2$ and $q^3$. ${\La}$ is
invariant under two different sets of gauge transformations, which
can be taken to be: \be
\delta_{\varepsilon}q^1=\varepsilon+\varepsilon q^2
\ddot{y},\;\;\delta_{\varepsilon}q^2=\varepsilon,\;\;
\delta_{\varepsilon}q^3=\varepsilon q^2 \ddot{y}
\label{gaugeex1}\ee and \be
\delta_{\eta}q^1=0,\;\;\delta_{\eta}q^2=\eta,\;\;\delta_{\eta}q^3=-\eta
\ee \label{gaugeex2}\no We can easily calculate the commutators of
the gauge transformations on the fields: \be
[\delta_{\varepsilon},\delta_{\eta}]\;q^1=\varepsilon
\eta\ddot{y},\;\;\;[\delta_{\varepsilon},\delta_{\eta}]
\;q^2=0,\;\;\;[\delta_{\varepsilon},\delta_{\eta}]\;q^3=
\varepsilon\eta\ddot{y}. \ee \no From (\ref{lagcoff}), the
equation of motion (eom) for $y$ is $\ddot{y}=0$. We see that the
algebra of the gauge transformations (\ref{gaugeex1}) and
(\ref{gaugeex2}) is
closed (in fact, abelian) only up to equations of motion, i.e., only on-shell. \\

\subsection{Reducible gauge theories}

The gauge transformations might also be ``reducible", i.e.,
dependent.  Consider the theory of an abelian $2$-form
$B_{\mu\nu}=-B_{\nu\mu}$.  The field strength is given by
$H_{\mu\nu\rho}=\partial_{\mu}B_{\nu\rho}+
\partial_{\nu}B_{\rho\mu}+\partial_{\rho}B_{\mu\nu}$. The Lagrangian reads:
\be {\La}=-\frac{1}{12}H_{\mu\nu\rho}H^{\mu\nu\rho} \label{lag2f}
\ee \no and is invariant under gauge transformations \be
\delta_{\Lambda}B_{\mu\nu}=\partial_{\mu}\Lambda_{\nu}-\partial_{\nu}\Lambda_{\mu}
\ee \no where $\Lambda$ is the gauge parameter. These
transformations vanish for a class of parameters
$\Lambda_{\mu}=\partial_{\mu}\epsilon$, meaning that the gauge
parameters are not all independent. Such gauge transformations are
called reducible, and the corresponding gauge theory is said to be
reducible. Two-forms define a natural generalization of
electromagnetism, $A_{\mu}\rightarrow B_{\mu\nu}$ and occur in
many models of unification; the main difference with
electromagnetism being the irreducibility of the latter.

\subsection{Reducibility on-shell}

The last feature that we want to illustrate is the possibility
that the reducibility of the gauge transformations holds only
on-shell. One can reformulate the previous free $2$-form model by
introducing an auxiliary field $A_{\mu}$. The Lagrangian is then
given by: \be
{\La}=\frac{1}{12}A_{\mu}\varepsilon^{\mu\nu\rho\sigma}
H_{\nu\rho\sigma}-\frac{1}{8}A_{\mu}A^{\mu}. \ee \no This
Lagrangian reduces to (\ref{lag2f}) by inserting the equation of
motion for $A_{\mu}$: \be
A_{\mu}=\frac{1}{3}\varepsilon^{\mu\nu\rho\sigma}H_{\nu\rho\sigma}.
\ee \no The gauge transformations are of the form:
\begin{eqnarray}
\delta_{\Lambda}B_{\mu\nu}&=&\partial_{\mu}\Lambda_{\nu}-
\partial_{\nu}\Lambda_{\mu} \\
\delta_{\Lambda}A_{\mu}&=&0.
\end{eqnarray}
\no We can introduce interactions by considering
Lie-algebra-valued fields $A_{\mu}=A_{\mu}^aT_a$,
$B_{\mu\nu}=B_{\mu\nu}^aT_a$ ($T_a$: generators of the gauge
group) and covariant derivatives instead of partial derivatives,
$\partial_{\mu} \rightarrow D_{\mu}$. This is the so-called
Freedman-Townsend model~\cite{Freedman:1980us}. \\

{}For the Freedman-Townsend model,  the gauge transformations
vanish for parameters $\Lambda_{\mu}=D_{\mu}\epsilon$.  But now,
this vanishing occurs only if the equations of motion are
satisfied. This is due to $[D_{\mu},D_{\nu}] \propto F_{\mu\nu}$,
and $F_{\mu\nu}=0$ is the eom for $B$. The theory is said in that
case to be reducible on-shell.

The antifield-BRST formalism is capable of handling all the gauge
structures described here, while the original methods were devised
only for off-shell closed, irreducible gauge algebras.  This wide
range of application of the antifield formalism is one of its main
virtues.

Remark: recent considerations on the structure of gauge
symmetries, including reducible ones, may be found in
\cite{Sardanashvily:2004ed,Bashkirov:2004nw}.

\section{Algebraic tools}
\setcounter{equation}{0} \label{algebraic}

BRST theory uses crucially cohomological ideas and tools. In the
following, some definitions are collected and a useful technique
for the computation of cohomologies is illustrated.

\subsection{Cohomology}

Let us consider a nilpotent linear operator $D$ of order $2$:
$D^2=0$. Because of this property, the image of $D$ is contained
in the kernel of $D$, Im $D$ $\subseteq$ Ker $D$. The cohomology
of the operator $D$ is defined as the following quotient space:
\be H(D)\equiv\frac{\mbox{Ker} D}{\mbox{Im} D}. \ee

\subsection{De Rham $d$}

As a familiar example, let us discuss the de Rham $d$-operator.
This will enable us to introduce further tools.

In a coordinate patch, a $p$-form is an object of the form \be
w=\frac{1}{p!}w_{i_1 \ldots i_p}\;dx^{i_1}\wedge \ldots \wedge
dx^{i_p} \ee \no where the coefficients $w_{i_1 \ldots i_p}$ are
totally antisymmetric functions of the coordinates and $\wedge$
refers to the exterior product. [Appropriate transition conditions
should hold in the overlap of two patches, but these will not be
discussed here.] The vector space of $p$-forms on $M$ is denoted
by $\Omega^p(M)$. The direct sum of $\Omega^p(M)$,
$p=0,\ldots,m\equiv \mbox{dim}\;M$ defines the space of all forms
on $M$: \be \Omega^*(M)\equiv \Omega^0(M) \oplus \ldots \oplus
\Omega^m(M) \ee \no The exterior derivative $\mbox{d}_p$ is a map
$\Omega^p(M) \rightarrow \Omega^{p+1}(M)$ whose action on a
$p$-form $w$ is defined by\footnote{The subscript $p$ is often not
written; the exterior derivative is then referred as d.} \be
\mbox{d}_p w=\frac{1}{p!} \frac{\partial w_{i_1 \ldots
i_p}}{\partial x^j}\;dx^j \wedge dx^{i_1}\wedge \ldots \wedge
dx^{i_p}. \ee \no Important properties of d are its nilpotency of
order two \be \mbox{d}^2=0
\;\;\;(\mbox{i.e.}\;\;\mbox{d}_{p+1}d_p=0) \label{prop1} \ee \no
and the fact that it is an odd derivative: \be \mbox{d}(w\wedge
\eta)=\mbox{d}w \wedge \eta + (-1)^p w \wedge \mbox{d}\eta
\label{prop2}. \ee \no The nilpotency can  easily be proved by
direct computation: \be \mbox{d}^2 w=\frac{1}{p!} \frac{\partial^2
w_{i_1 \ldots i_p}}{\partial x^k\partial x^j} \;dx^k \wedge dx^j
\wedge dx^{i_1}\wedge \ldots \wedge dx^{i_p}. \ee \no This
expression clearly vanishes since the coefficients are symmetric
in $k,j$ while $dx^k \wedge dx^j$ is antisymmetric. Therefore, Im
$\mbox{d}_p$ $\subset$ Ker $\mbox{d}_{p+1}$.

An element of Ker $\mbox{d}_{p}$ is said to be ``closed" or ``a
cocycle" (of d), d$\alpha=0$. An ``exact form" or ``coboundary"
(of d) lives in Im $\mbox{d}_{p-1}$; it is thus such that
$\alpha=\mbox{d}\beta$, for some $(p-1)$-form $\beta$. The $p$th
de Rham cohomology group is defined as: \be
H^p(\mbox{d})=\frac{\mbox{Ker}\; \mbox{d}_p}{\mbox{Im}\;
\mbox{d}_{p-1}}. \ee \no Another important operation is the
interior product $\mbox{i}_X$: $\Omega^p(M) \rightarrow
\Omega^{p-1}(M)$, where $X=X^{j}\partial/\partial x^{j}$. The
action of $\mbox{i}_X$ on a $p$-form $w$ reads: \be \mbox{i}_X
w=\frac{1}{(p-1)!}X^{j}   w_{j i_2 \ldots i_p} \;dx^{i_2}\wedge
\ldots \wedge dx^{i_p}. \ee \no The interior product also satisfy
(\ref{prop1}) and (\ref{prop2}) as well.

\subsection{Poincar\'{e} lemma}
A useful tool for computing cohomologies is given by contracting
homotopies.  We illustrate the techniques by computing the
cohomology of $d$ in a special case.

Let $M$ be $\mathbb{R}^n$, and $w$ a closed $p$-form on M, $p>0$.
The Poincar\'{e} lemma states that all closed forms in degree $>0$
are exact\footnote{For $M \not=\mathbb{R}^n$, closed forms might
not be globally exact (although they are locally so). The
Poincar\'{e} lemma fails in such a case.}. More precisely:
\begin{eqnarray}
H^p(\mbox{d})&=&0 \;\;\;\;p>0 \\
H^p(\mbox{d}) & \simeq & \mathbb{R} \;\;\;p=0. \nonumber
\end{eqnarray}
\no {\bf{Proof}}: \\
\no Assume the coefficients $w_{i_1 \ldots i_p}$ to be polynomial
in $x^i$ (this restriction is not necessary and is made only to
simplify the discussion). Let: \be
\mbox{d}=dx^i\frac{\partial}{\partial
x^i},\;\;\;i_x=x^i\frac{\partial^L}{\partial (dx^i)}\; (\equiv
i_X,\;X^{j}=x^{j}). \ee \no Define the counting operator $N$ as the
combination\footnote{It is the Lie derivative of a form along $X$,
$N w = {\mathcal{L}}_Xw$.} $N=\mbox{d}i_X+i_X\mbox{d}$. It is such
that $Nx^i=x^i$, $Ndx^i=dx^i$. For a general form $\alpha$ we have
$N \alpha=k\alpha$ with $k$ the total polynomial degree (in $x^i$
and $dx^i$). It then follows that for:
\begin{itemize} \item  $ k \neq 0$:  $$\alpha=\frac{k}{k}\alpha=N\left(\frac{1}{k}\alpha\right)=
(\mbox{d}i_x+i_x\mbox{d})\left(\frac{1}{k}\alpha\right).$$ Thus, if
$\mbox{d}\alpha=0$ then $\alpha=\mbox{d}(i_{x}
\left(\frac{1}{k}\alpha\right))$. \item $ k = 0$: $\mbox{Im}\;
\mbox{d}_{-1}$ has no meaning, there is no such a thing as a
$(-1)$-form. We can say that $\Omega^{-1}(M)$ is empty and
$H^0(\mbox{d})=\mbox{Ker}\; \mbox{d}_0$. So constants are the only
members of the cohomology.
\end{itemize}

\no This proves the Poincar\'e lemma.

\subsection{Local functions}

A local function $f$ is a smooth function of the spacetime
coordinates, the field variables and their respective derivatives
up to a finite order,
$f=f(x,[\varphi])=f(x^{\mu},\varphi^{i},\partial_{\mu}\varphi^{i}, \cdots,
\partial_{\mu_1 \cdots \mu_k}\varphi^{i})$. In field theory local
functions are usually polynomial in the derivatives. The following
discussion is however more general than that, and it remains valid
in the case of arbitrary smooth local functions. \\

\no The Euler-Lagrange derivative $\frac{\delta }{\delta\varphi^i}$ of a local function $f$ is defined by
\be
\frac{\delta f}{\delta\varphi^i}=\sum_{k\geq 0}(-)^k\partial_{\mu_1}\ldots
\partial_{\mu_k}\frac{\partial f}{\partial(\partial_{\mu_1}\dots \partial_{\mu_k}\varphi^i)}
\ee
\no (with $\partial_{\mu_1}\dots \partial_{\mu_k}\varphi^i)$ the last derivative of $\varphi$ occurring in
$f$). \\

\no {\bf{Theorem :}} A local function is a total derivative iff it
has vanishing Euler-Lagrange derivatives with respect to all
fields: \be f=\partial_{\mu}j^{\mu}\;\;\;\Leftrightarrow
\;\;\;\frac{\delta f}{\delta\varphi^i}=0 \;\; \;\; \forall
\varphi^i. \ee \no A proof of this theorem will not be given here
but can be found for instance
in~\cite{Brandt:1989gy,Dubois-Violette:1992ye,Barnich:2000zw}
(see also references given in \cite{Barnich:2000zw}). \\

\subsection{Local differential forms}
Local $p$-forms are differential forms whose coefficients are local functions:
\be
w = \frac{1}{p!}w_{i_1 \ldots i_p}(x,[\varphi])\;dx^{i_1}\wedge \ldots \wedge dx^{i_p}
\ee
\no Consider a local $n$-form $w=fd^nx$ in ${\mathbb{R}}^n$; $w$ is trivially closed.
It is further exact iff the function $f$ is a total derivative:
\be
w=\mbox{d}\alpha \;\;\;\Leftrightarrow \;\;\; f=\partial_{\mu}j^{\mu}\;\;\;
\Leftrightarrow \;\;\;\frac{\delta f}{\delta\varphi^i}=0 \;\; \;\; \forall \varphi^i \label{exactform}
\ee

\subsection{Algebraic Poincar\'{e} lemma}
The algebraic Poincar\'{e} lemma gives the cohomology of d in the algebra of local forms:
\begin{eqnarray}
p=n: \;\;\;H^p(\mbox{d})&\not=&0 \;\; \;\;\mbox{and characterized
above} \nonumber \\
0<p<n: \;\;\;H^p(\mbox{d})&=&0 \;\;\;\; \\
p=0:\;\;\; H^p(\mbox{d}) & \simeq & \mathbb{R}.  \nonumber
\end{eqnarray}
\no The distinguishing feature compared with the Poincar\'{e}
lemma for ordinary exterior forms not depending on local fields is
the appearance of a non-vanishing cohomology at $p=n$; non trivial
local $n$-forms $w = fd^nx$ are such that at least one of the
derivatives $\delta f/\delta\varphi^i$ is not identically zero. A
proof of the algebraic Poincar\'{e} lemma in the case of
polynomial dependence on derivatives can be found
in~\cite{Barnich:2000zw}; for a more general proof see
\cite{Brandt:1989gy,Dubois-Violette:1992ye}. \\

\no {\bf{Example}}. Consider local $1$-forms in $\mathbb{R}^1$,
$w={\La}(t,q,\dot{q},\ddot{q},\ldots)\;dt$. For example: \be
w_1={\La}_1\;dt=\frac{1}{2}\dot{q}^2 dt,\;\;\;
w_2={\La}_2\;dt=\dot{q}q dt. \ee \no These are obviously closed,
but are they exact? We work out the Euler-Lagrange derivatives of
${\La}_{1,2}$: \be \frac{\delta {\La}_1}{\delta
q}=-\ddot{q},\;\;\;\frac{\delta {\La}_2}{\delta q}=0. \ee \no
Therefore, from (\ref{exactform}), $w_2$ is an exact form (and
indeed, $w_2 = \mbox{d}\left(\frac{1}{2} q^2\right)$), while $w_1$
cannot be written as the exterior derivative of a local function.
$H^1(\mbox{d})$ is clearly non trivial.

\section{BRST construction}
\setcounter{equation}{0} \label{BRSTconstruction}

We stated in the introduction that gauge invariance is lost after
the necessary gauge fixing. The central idea of the BRST
construction~\cite{Becchi:1975nq, Tyutin:1975qk} is to replace the
original gauge symmetry by a rigid symmetry, the BRST symmetry $s$,
which is still present even after one has fixed the gauge. This is
achieved by introducing extra fields in the theory: the ghost fields
and the conjugate antifields. The operator $s$ acts on the enlarged
space of fields, ghosts and antifields. An extended action involving
all these variables can be constructed in such a way that it is BRST
invariant. The operator $s$ is called the BRST differential and it
is nilpotent: $s^2=0$. Therefore, cohomological groups $H^k(s)$ can
be constructed. The BRST differential fulfills: \be
H^0(s)={\mbox{Gauge invariant functions (``Observables'')}} \ee \no
In this way we recover the gauge symmetry.  This is BRST theory in a
nutshell.
These important statements will now be discussed in more detail. \\

\subsection{Master equation}

Consider a gauge theory of fields $\varphi^i$ described by a
classical action $S_0(\varphi^i)$ on a manifold $M$. The equations
of motion constrain the fields  to a submanifold, denoted
$\Sigma$. The action is invariant under gauge
transformations\footnote{We use De Witt's condensed notation; see
Appendix I.} \be
\delta_{\varepsilon}\varphi^i=R^i_{\alpha}\varepsilon^{\alpha}.
\label{gtransg} \ee \no Assume the theory to be (on-shell)
reducible, with no reducibility on the reducibility functions. In
such a case there are relations among the gauge parameters but no
relations among the relations. The relations among the gauge
parameters can be written as: \be
Z^{\alpha}_{\Delta}R^i_{\alpha}=C_{\Delta}^{ij}\frac{\delta
S_0}{\delta \varphi^j }. \ee \no We proceed as follows. For each
commuting (anticommuting) gauge parameter $\varepsilon^{\alpha}$
one introduces a fermionic (bosonic) ghost variable $c^{\alpha}$.
We also introduce ghosts of ghosts $c^{\Delta}$, one for each
(independent) reducibility identity of the theory. The set of
original fields, ghosts and ghosts of ghosts are collectively
denoted as $\Phi^A$. We double now the configuration space by
considering conjugate fields w.r.t.\ each of the $\Phi$' s: the
anti-fields $\Phi^*_A$. They are postulated to have opposite
(Grassmann) parity. Gradings are assigned to the
various fields as displayed in table \ref{tab1}.\\

\begin{table}[h]
\begin{center}
\begin{tabular}{|r|c|c|c|}

\hline
& $\puregh$& $\antifd$& $\gh$ \\
\hline
$\varphi^i$ &$0$&$0$&$0$ \\

$c^{\alpha}$ &$1$&$0$&$1$ \\
$c^{\Delta}$ &$2$&$0$&$2$ \\
$\varphi_i^*$  &$0$&$1$&$-1\;\;$ \\
$c_{\alpha}^*$ &$0$&$2$&$-2\;\;$ \\
$c_{\Delta}^*$ &$0$&$3$&$-3\;\;$ \\
\hline

\end{tabular}
\caption{Pure ghost number, antifield number and $\gh \equiv
\puregh-\antifd$ (``total ghost number"), for the different field
types.}\label{tab1}
\end{center}
\end{table}

\no We define the antibracket\footnote{Properties of anti-brackets
are listed in Appendix II.} of two functionals
$F(\Phi^A,\Phi_A^*)$, $G(\Phi^A,\Phi_A^*)$ by: \be
(F,G)=\frac{\delta^R F}{\delta\Phi^A}\frac{\delta^L
G}{\delta\Phi_A^*}- \frac{\delta^R
F}{\delta\Phi_A^*}\frac{\delta^L G}{\delta\Phi^A}. \ee \no Here,
$L$ (respectively $R$) refers to the standard left (respectively
right) derivative.  These are related as \be \frac{\delta^R F}
{\delta (\mbox{field}) } \equiv (-1)^{\varepsilon_{\mbox{\tiny
field}}(\varepsilon_F+1)}\frac{\delta^L F} {\delta
(\mbox{field})}, \ee \no where $\varepsilon$ denotes the
Grassmann-parity. The BRST transformation of any functional
$F(\Phi^A,\Phi_A^*)$ can be written in terms of antibrackets: \be
s F=(S,F). \ee \no The generating function $S(\Phi^A,\Phi_A^*)$ of
the BRST transformation is sometimes called the generalized
action. The BRST transformation is nilpotent of order two; this is
reflected in the (classical) master equation: \be \fbox{(S,S)=0}
\ee \no The solution to the master equation is unique up to
canonical transformations. It can be constructed in a sequential
form\footnote{There is however no guarantee for this sequence to
be finite!}: \ba
S&=&S_0+S_1+S_2+ \ldots \;\;\; \nonumber \\
 \\
S_0 \equiv \mbox{classical
action},\;\;\;S_1&=&\varphi_i^*R^i_{\alpha}c^{\alpha},\;\;
S_2=c_{\alpha}^*Z^{\alpha}_{\Delta}c^{\Delta}+ \ldots \nonumber
\ea \no The proof of this statement (including the reducible case)
can be found in
\cite{Batalin:1985qj,Fisch:1989rp,Henneaux:1989jq,Henneaux:1992ig}
and references therein. Locality of $S$ under general conditions
is established in \cite{Henneaux:1990rx}.

The solution $S$ of the master equation is key to the
BRST-antifield formalism. It can be written down explicitly for
the Yang-Mills theory and the abelian 2-form model introduced earlier: \\

\no {\bf{Yang-Mills.}}
\be
S= -\frac{1}{4}\int d^nx \;F_{\mu\nu}^aF^{\mu\nu}_a+\int d^nx \;A^{*\mu}_{\;\;a}
D_{\mu}c^a+\frac{1}{2}\int d^nx \;c^*_a f^a_{\;\;bc}c^bc^c
\ee

\no
{\bf{Abelian $\mathbf{2}$-form.}}
\be
S= -\frac{1}{12}\int d^nx \;H_{\mu\nu\rho}H^{\mu\nu\rho}+\int d^nx \;
B^{*\mu\nu}(\partial_{\mu}c_{\nu}-\partial_{\nu}c_{\mu})+\int d^nx
\;c^{*\mu}\partial_{\mu}c
\ee
\\
In those cases, the solution $S$ of the master equation is linear in
the antifields. {}For gauge systems with an ``open algebra" (i.e.,
for which the gauge transformations close only on-shell), or for
on-shell reducible gauge theories, the solution of the master
equation is more complicated.  It contains terms that are indeed non
linear in the antifields. These terms are essential for getting the
correct gauge fixed action below. Without them, one would not derive
the correct Feynman rules leading to gauge-independent amplitudes.
\\
{\vskip.1cm}

\no
{\bf{Exercises.}} \\

\no
1.) Consider a nilpotent operator $\Omega$ of order $N$ (i.e. $\Omega^N\equiv0$). \\

\indent (i) prove that the only eigenvalue of $\Omega$ is zero. \\
\indent (ii) for $N=2$, analyze the cohomology of $\Omega$
 in terms of its Jordan decomposition. \\
\indent (iii) for $N=3$, Im $\Omega^2$ $\subset$ Ker $\Omega$. The
corresponding cohomologies are defined as \be
H_{(1)}(\Omega)\equiv\frac{\mbox{Ker} \Omega}{\mbox{Im} \Omega^2},
\;\;\;H_{(2)}(\Omega)\equiv\frac{\mbox{Ker} \Omega^2}{\mbox{Im}
\Omega} \nonumber \ee Calculate these. \\
 \no Hint: The Jordan
decomposition of a matrix is a block-diagonal form. Each such
block, called Jordan block, has on its diagonal always the same
eigenvalue and 1 in the upper secondary diagonal.\\

\noindent 2.) Prove that
P$(\varphi^i,\partial_{\mu}\varphi^i,\ldots,
\partial_{\mu_1 \ldots\mu_k}\varphi^i)\;d^nx$ is exact iff $\frac{\delta P}
{\delta\varphi^i} =0$, where $\frac{\delta}{\delta\varphi^i}$ is the Euler-Lagrange derivative. \\

\no
Hint: Relate $N=\varphi^i\frac{\partial}{\partial\varphi^i}+(\partial_{\mu}\varphi^i)
\frac{\partial}{\partial(\partial_{\mu}\varphi^i)}+\ldots \; \; $  to $\frac{\delta P}{\delta\varphi^i}$ \\

\noindent 3.) Write explicitly $R^i_{\alpha}$, $Z_{\Delta}^{\alpha}$, $C_{\Delta}^{ij}$
for the Freedman-Townsend model. \\

\noindent 4.) Write Noether's identities (see Appendix I) for
Yang-Mills,
gravity and an abelian $2$-form gauge theory. \\

\noindent 5.) Check the properties of anti-brackets given in Appendix II. \\

\noindent 6.) Consider an irreducible gauge theory with gauge
transformations closing off-shell and forming a group. The
solution of the master equation is given by

\be
S=S_0 +\varphi_i^*R^i_{\alpha}c^{\alpha}+\frac{1}{2}c^*_{\alpha}f^{\alpha}_{\;\;\beta\gamma}
c^{\beta}c^{\gamma}
\ee
\no
where $f^{\alpha}_{\;\;\beta\gamma}$ are the structure constants of the gauge group.
Verify that $S$ satisfies the master equation $(S,S)=0$. \\

\noindent 7.) Define the operator $\Delta$ as
\be
\Delta F=(-1)^{\varepsilon_A}\frac{\delta^L}{\delta\phi^A}\frac{\delta^LF}
{\delta\phi_A^*}.\label{exedelta}
\ee
\no Prove the following statements: \\

\indent (i) $\varepsilon(\Delta)=1$ \\
\indent (ii) $\Delta^2=0$ \\
\indent (iii) gh $\Delta=1$ \\
\indent (iv) $\Delta (\alpha, \beta)=(\Delta\alpha,\beta)-
(\alpha,\Delta\beta)(-1)^{\varepsilon_{\alpha}}$ \\
\indent (v) $\Delta (\alpha \beta)=(\Delta\alpha)\beta+
(-1)^{\varepsilon_{\alpha}}\alpha(\Delta\beta)+(-1)^{\varepsilon_{\alpha}}(\alpha,\beta)$ \\

\no Verify that the superjacobian for the change of variables
$\phi^A \rightarrow \phi^{'A}+({\mu} S,\phi^A)$
is $1-({\Delta} S)\mu$, where $\mu$ is a fermionic constant. \\

\section{Observables}
\setcounter{equation}{0} \label{observables}

It is now time to substantiate the claim \be
H^0(s)\simeq\mathrm{Observables}\label{claimobs} \ee where, as we
have just seen, the BRST differential is given by $sF = (S,F)$. In
particular, \bea s\Phi^A&=&(S,\Phi^A)=-\frac{\delta^R
S}{\delta\Phi^*_A}=
-\sum_k\frac{\delta^R S_k}{\delta\Phi^*_A}\label{fieldtrans}\\
s\Phi^*_A&=&(S,\Phi^*_A)=\frac{\delta^R
S}{\delta\Phi^A}=\sum_k\frac{\delta^R
S_k}{\delta\Phi^A}.\label{afieldtrans} \eea Note that the ghost
number of $S$ is 0, the ghost number of the BRST transformation is
1 as well as the one of the antibracket, so that the gradings of
both sides of the equation $sF=(S,F)$ match. We shall actually not
provide the detailed proof of (\ref{claimobs}) here, but instead,
we give only the key ingredients that underlie it,
referring again to
\cite{Fisch:1989rp,Henneaux:1989jq,Henneaux:1992ig} for more
information.

To that end, we expand the BRST transformations of all the
variables according to the antifield number, as in
\cite{Fisch:1989dq,Fisch:1989rp,Henneaux:1989jq,Henneaux:1992ig}.
So one has, \bea
S&=&\sum_{k\ge 0} S_k\\
s&=&\delta+\gamma+\sum_{i>0} s_i.\label{sexpand} \eea \noindent
The first term $\delta$ has antifield number -1 and is called the
Koszul-Tate differential, the second term $\gamma$ has antifield
number 0 and is called the longitudinal differential, and the next
terms $s_i$ have antifield number $i$.  Although the expansion
stops at $\delta + \gamma$ for Yang-Mills (as it follows from the
solution of the master equation given above), higher order terms
are present for gauge theories with an open algebra, or on-shell
reducible theories\footnote{The expansion of $s$ is connected to
spectral sequences, which will not be discussed in detail here
\cite{Henneaux:1992ig}.}.

Explicitly, one finds for the Koszul-Tate differential
$\delta\varphi^i=0$, as there is no field operator of anti-field
number -1 and $\delta\varphi_i^*=\delta S_0/\delta\varphi^i$ . For
the longitudinal differential it follows in the same way that \be
\gamma\varphi^i=R_\alpha^i c^\alpha\label{gammaphi}. \ee Observe
also that \be
0=s^2=\delta^2+\{\delta,\gamma\}+(\gamma^2+\{\delta,s_1\})+...\label{specseq}
\ee \noindent In this equation, each term has to vanish
separately, as each term is of different antifield number.

Let $A$ be a BRST-closed function(al), $sA = 0$.  We must compute
the equivalence class \be A \sim A+sB.\label{genobj} \ee \noindent
Since we are dealing with observables, the only relevant operators
are of of ghost number 0, thus $\gh A=0$ and $\gh B=-1$. The
latter can only be satisfied, if $B$ contains at least one
anti-field. Expanding $A$ and $B$ in antifield number yields \bea
A=\sum_{k\ge 0} A_k=A_0+A_1+A_2+...\\
B=\sum_{k\ge 1} B_k=B_1+B_2+B_3+... \eea \noindent  Acting with
$s$ on $A$ using the expansion \pref{sexpand} then gives \be
(\delta+\gamma+...)(A_0+A_1+...)=(\gamma A_0+\delta A_1)+..., \ee
\noindent where the term in parentheses collects all antifield
number zero contributions. The condition $sA=0$ implies that this
term must vanish on its own and thus $\gamma A_0=-\delta A_1$.
{}Furthermore,one finds \be
A+sB=A+(\delta+\gamma+...)B=(A_0+\delta B_1)+... \ee \noindent
where the last term in parentheses is again the antifield zero
contribution. Using \pref{afieldtrans}, \be \delta
B_1=\frac{\delta B_1}{\delta\Phi^*}\frac{\delta
S_0}{\delta\Phi^i}, \ee \noindent we see that the second term of
the antifield zero contribution in $A + sB$ vanishes when the
equations of motions are fulfilled, i.e.\ on-shell. A similar
property holds for $\delta A_1$.  There is therefore a clear
connection of $\delta$ to the dynamics and the equations of
motions. Note that the ``on-shell functions" can be viewed as the
equivalence classes of functions on $M$ identified when they
coincide on $\Sigma$, i.e.,  $C^\infty(\Sigma)=C^\infty(M)/{\cal
N}$, where the ideal ${\cal N}$ contains all the functions that
vanish on-shell.

From (\ref{gammaphi}) and the fact that $\gamma$ is a derivation,
one gets for\footnote{$A_0$ can only depend on $\phi_i$, as it has
pure ghost and antifield number 0.}$\gamma A_0$ \be \gamma
A_0=\frac{\delta A_0}{\delta\varphi^i}R_\alpha^i c^\alpha. \ee
\noindent As this is a gauge transformation (with gauge parameters
replaced by the ghosts), $\gamma A_0$ vanishes if $A_0$ is
gauge-invariant. The longitudinal differential is associated with
gauge transformations. We thus see that a necessary condition for
$A$ to be BRST-closed is that its first term $A_0$ be
gauge-invariant on-shell.  And furthermore, two such $A_0$'s are
equivalent when they coincide on-shell.

It turns out that the condition on $A_0$ is also sufficient for
$A$ to be BRST-closed, in the sense that given an $A_0$ that is
gauge-invariant on-shell, one can complete it by terms $A_1$,
$A_2$ ... of higher antifield number so that $sA=0$.

To summarize: the term that determines the cohomological class of
a BRST cocycle is the first term $A_0$.  This term must be an
observable, in that it must be gauge invariant on-shell. We can
therefore conclude that $H^0(s)$ captures indeed the concept of
observables.  The differential $\delta$ reduces from the manifold
$M$ to the on-shell manifold $\Sigma$ and $\gamma$ further to
$\Sigma/G$, the set of all gauge-invariant functions, where $G$ is
the set of all gauge orbits.

\section{Path Integral and Gauge-fixing}
\setcounter{equation}{0} \label{gauge-fixing}

We first consider the Yang-Mills case. To perform actual
path-integral calculations, it is necessary to gauge-fix the
theory. To perform this task, it is convenient to add additional
fields, the anti-ghost $\bar c_a$ and auxiliary fields, the
Nakanishi-Lautrup fields $b_a$. They transform as $s\bar c_a\sim
b_a$ and $s b_a=0$. We take $\bar c_a$ and $b_a$ to have ghost
number -1 and 0, respectively. The corresponding antifields $\bar
c^{a*}$ and $b^{a*}$ have thus ghost number 0 and -1,
respectively. Furthermore a contracting homotopy argument similar
to the one given above for the Poincar\'e lemma shows that the
counting operator of $\bar c_a$, $b_a$ and their conjugate
antifields is BRST exact. Hence the cohomology is not altered by
the introduction of these new variables.  In particular, the set
of observables is not affected. The solution of the master
equation with the new variables included reads, for Yang-Mills
theory  \bea
S&=&-\frac{1}{4}\int d^nx F_\mn^a F^\mn_a+\int d^nx A_a^{\mu*}D_\mu c^a\nonumber\\
&&+\frac{1}{2}\int d^nx c_a^* f^a_{\;\;bc} c^b c^c-i\int d^nx\bar c^{*a}b_a.
\eea
\noindent The last term is called the non-minimal part.

\vspace{.2cm} \noindent {\bf Theorem}

\noindent  {\it The generating functional \be {\cal Z} = \int{\cal
D}\Phi^A\exp\left(\frac{i}{\hbar}S_\psi[\Phi^A]\right),
\label{genfuct}\ee \noindent does not depend on the choice of
$\psi$. Here, $\psi$ is called the gauge-fixing fermion, and has
Grassmann-parity 1 (hence its name) and ghost number -1. In
(\ref{genfuct}), the notation \bea
\Phi^a&=&(A_\mu^a,c^a,\bar c^a,b_a)\\
\Phi^*_a&=&(A_\mu^{a*},c^{a*},\bar c^{a*},b^{a*}), \eea has been
used and the ``gauge-fixed action" $S_\psi[\Phi^A]$ is given by
\be
S_\psi[\Phi^A]=S\left[\Phi^A,\Phi^*_A=\frac{\delta\psi}{\delta\Phi^A}\right].
\ee } \hspace{.2cm}

\noindent This theorem is proved in section
\ref{quantummaster} below.
\hspace{.2cm}

Before turning to the proof, we want to illustrate formula
(\ref{genfuct}) by showing how one can choose the gauge-fixing
fermion $\psi$ to reproduce familiar expressions for the path
integral of the Yang-Mills field.   A possible choice, which leads
to non-degenerate propagators for all fields and ghosts, is given
by \be \psi=i\int d^nx\bar c^a\left({\cal
F}^a+\frac{\alpha}{2}b^a\right), \ee \noindent where ${\cal F}^a$
is the gauge condition, e.g.\ ${\cal F}^a=\pd^\mu A^{a}_{\mu}$ for
covariant gauges and $\alpha$ is the gauge parameter. This leads
to \bea
\bar c^{a*}&=\frac{\delta\psi}{\delta\bar c_a}=&i\left({\cal F}^a+\frac{\alpha}{2}b^a\right)\\
A_{a\mu}^*&=\frac{\delta\psi}{\delta A_\mu^a}=&-i\pdm\bar c_a\\
b^{a*}&=\frac{\delta\psi}{\delta b_a}=&i\frac{\alpha}{2}\bar c^a\\
c_{a}^*&=\frac{\delta\psi}{\delta c^a}=&0. \eea \noindent One then
gets the familiar gauge-fixed Yang-Mills action \bea
&&S_\psi[A_\mu^a,c^a,\bar c_a,b_a]\nonumber\\
&&=\int d^nx\left(-\frac{1}{4}F_\mn^a F^\mn_a-i\pd^\mu\bar c_a D_\mu
c^a+\left({\cal F}^a+\frac{\alpha}{2}b^a\right)b_a\right), \eea
which is usually obtained by the Fadeev-Popov procedure
\cite{Faddeev:1967fc}.  The conventional Landau gauge is recovered
by setting $\alpha=0$. As the resulting path integral can be written
as \be \int{\cal D}[A_\mu^a\bar c^a c^a]\delta(\pd^\mu
A_\mu^a)e^{\left(\frac{i}{\hbar}S_{gf}[A_\mu^a,\bar c^a,
c^a]\right)}, \ee \noindent the transversality of the gauge boson is
directly implemented. Landau gauge is thus called a strict gauge. An
example of a non-strict gauge is Feynman gauge, $\alpha=1$, in which
case the equations of motion do not imply ${\cal F}^a=0$ but yield
instead $b^a\sim{\cal F}^a$.

The choice of the gauge-fixing fermion is not unique. One can add
to $\psi$ the term $\psi\sim\bar c_a\bar c_b c^c$, which yields
quartic ghost couplings. Quartic ghost renormalizations may even
be needed without such explicit terms, e.g.\ when using ${\cal
F}^a=\pd^\mu A_\mu^a+d^a_{\;\;bc}A_\mu^b A^{\mu c}$, where
$d^a_{\;\;bc}$ is a symmetric tensor in color space (see
\cite{Zinn-Justin:1984dt,Weinberg:1996kr}).

By appropriately choosing the gauge fixing fermion, one can reduce
the path integral to an expression that involves only the physical
(transverse) degrees of freedom and which is manifestly unitary in
the physical subspace (equal to the reduced phase space path
integral). Independence on the choice of $\psi$ (still to be
proved) guarantees then that the expression (\ref{genfuct}) is
correct.  We shall not demonstrate here the equivalence of
(\ref{genfuct}) with the reduced phase space path integral.  The
reader may find a discussion of that point in
\cite{Henneaux:1992ig}.

As a final point, we note that in order for (\ref{genfuct}) to be
indeed independent on the choice of $\psi$,  it is necessary that
the measure be BRST invariant. This can be investigated using the
operator $\Delta$, already defined in the exercises in
\pref{exedelta} as \be
\Delta=(-1)^{\epsilon_A}\frac{\delta^L}{\delta\Phi^A}\frac{\delta^L}{\delta\Phi^*_A}.\label{defdelta}
\ee \noindent The BRST transformation can be written as \be
\Phi^A\to\Phi^{A'}=\Phi^A+(\mu
S,\Phi^A)=\Phi^A-(\Phi^A,S)\mu=\Phi^A-\frac{\delta^L
S}{\delta\Phi_A^*}\mu, \ee \noindent where $\mu$ is a constant,
anti-commuting parameter. The Jacobian of this transformation is
given by \be
J_{AB}=\frac{\delta^L\Phi^{A'}}{\delta\Phi^B}=\delta^{AB}-\frac{\delta^L}{\delta\Phi^B}\frac{\delta^L
S}{\delta\Phi_A^*}\mu. \ee \noindent [As the Jacobian involves
commuting and anti-commuting fields, the Jacobian ``determinant"
is actually a super-determinant.] Therefore the measure transforms
as \be {\cal D}\Phi^A\to\sdet J\;{\cal D}\Phi^{A'}. \ee \noindent
For an infinitesimal transformation, the super-determinant can be
approximated by the super-trace \be \sdet
J\approx1+\str\left(-\frac{\delta^L\delta^L
S}{\delta\Phi_B\delta\Phi^{A*}}\mu\right)=1-(-1)^{\epsilon_A}\frac{\delta^L\delta^L
S}{\delta\Phi_A\delta\Phi^{A*}}\mu=1+\Delta S. \ee \noindent It
follows that the measure is BRST-invariant iff $\Delta S=0$. The
property $\Delta S=0$ can be shown by explicit calculation for
pure Yang-Mills theory \cite{DeJonghe:1992um}. The more general
case will be treated in the last section.  Further interesting
properties of $\Delta$ and of the formalism are developed in \cite{Batalin:1984ss}.\\

\no{\bf{Exercises.}} \\

\no 8.) For Yang-Mills theory in 4 dimensions, compute the
dimensionality of all fields. Is there any freedom? What is the
most general gauge fixing fermion $\psi$ of mass dimension 3? What
are the restrictions on the gauge-fixing fermion, if the action is
required to be invariant under the transformation $\bar c^a\to\bar
c^a+\epsilon^a$?

\no 9.) (a) Show that for a functional $W$ with $\gh W=0$ and Grassmann-parity 0
\be
\Delta e^{\frac{i}{\hbar} W}=0\Leftrightarrow\frac{1}{2}(W,W)-i\hbar\Delta W=0,\label{exew}
\ee
\noindent where $\Delta$ is the operator defined by \pref{defdelta}.\\
\noindent (b) Define the operator $\sigma$ as \be
\sigma\alpha\equiv(W,\alpha)-i\hbar\alpha, \ee \noindent where $W$
satisfies \pref{exew}. Then show that \be
\sigma\alpha=0\Leftrightarrow\Delta(\alpha e^{\frac{i}{\hbar}
W})=0. \ee
\noindent (c) Show that $\sigma$ is nilpotent, $\sigma^2=0$.\\
\noindent (d) Show that if $\alpha=\sigma\beta$, then $\alpha\exp{iW/\hbar}=\Delta$(something).

\no 10.) Write explicitly the action of $s$, $\delta$, and
$\gamma$ on all fields and antifields for Yang-Mills theory.

\no 11.) For Yang-Mills theory, consider $H(\gamma)$ in the space
of polynomials in the ghost fields $c^a$, i.e. $a_0+a_a c^a+a_{ab}
c^a c^b+...$. Compute $H^0(\gamma)$ and $H^1(\gamma)$. Show in
particular that $H^2(\gamma)$ parameterizes the non-trivial
central extensions $h_{ab}$, i.e., non trivial modifications of
the algebra of the form $[X_a,X_b]=f^a_{\;\;bc}X_a+h_{ab}1$.

\section{Beyond Yang-Mills}
\setcounter{equation}{0} \label{beyond}

The results of the previous section generalize straightforwardly
to gauge theories other than Yang-Mills. The solution of the
master equation $S$ is of the form \be S=S_0+\phi_i^*R_\alpha^i
c^\alpha+..., \ee \noindent where $S_0$ is the classical action.
The second term is uniquely determined by the gauge transformation
\pref{gtransg}, and all further terms depend on the specific
theory. While the expansion of the solution of the master equation
stops at antifield number one in the Yang-Mills case, one gets
higher order terms in the case of open gauge systems.  It is again
often convenient to extend to the non-minimal sector by
introducing $\bar c_\alpha$ and $b_\alpha$ and the corresponding
antifields in a similar manner to what has been done in the case
of Yang-Mills theory.

Assuming again $\Delta S=0$, the quantized theory follows from a
path integral \be {\cal Z}=\int{\cal
D}\Phi^A\exp\left(\frac{i}{\hbar} S_\psi[\Phi^A]\right). \ee
\noindent $S_\psi$ is the solution $S$ of the master equation
$(S,S) = 0$, in which the antifields have been eliminated by use
of the gauge-fixing fermion $\psi$ as before, \be
S_\psi[\Phi^A]=S_\psi\left[\Phi^A,\Phi_A^*=\frac{\delta\psi}{\delta\Phi^A}\right].\label{replaceds}
\ee The gauge-fixing fermion has again odd Grassmann-parity and
ghost number -1. It is in general given by a local expression \be
\psi=\int
d^nx\chi(\Phi^A,\pdm\Phi^A,...,\pd_{\mu_1}...\pd_{\mu_k}\Phi^A).
\ee \noindent {}For theories with an open algebra, the terms
quadratic in the antifields will lead to quartic (or higher)
ghost-antighost vertices in the gauge-fixed action.  While these
terms are a gauge-dependent option in the Yang-Mills case, they
have an unavoidable character (in relativistic gauges) for open
gauge algebras.  These terms, which follow directly from the
general construction of the gauge-fixed action, cannot be obtained
through the exponentiation of a determinant, since this procedure
always produces an expression which is quadratic in the ghosts.

A useful concept is that of gauge-fixed BRST transformation, which
is what $s$ becomes after gauge-fixing.  It is denoted by $s_\psi$
and defined as \be
s_\psi\Phi^A=(s\Phi^A)|_{\Phi_A^*=\delta\psi/\delta\Phi^A}. \ee
\noindent Note that $s_\psi^2=0$ is in general only valid on
(gauge-fixed) shell, i.e.\ for field configurations satisfying
$\delta S_\psi/\delta\Phi^A=0$.

If $S$ is linear in the antifields, i.e.\ if the gauge algebra
closes off-shell, $s_\psi\Phi^A=s\Phi^A$. This follows directly from
\be s\Phi^A=(S,\Phi^A)=-\frac{\delta^R
S}{\delta\Phi_A^*},\label{linbrsttrans} \ee \noindent which is
independent of the antifields, if $S$ depends only linearly on the
antifields. In that case $s_\psi^2=0$ even off-shell provided one
keeps all the variables.  This is the case in Yang-Mills theory. [In
Yang-Mills theory, one often eliminates the auxiliary fields $b_a$
by means of their own equations of motion. One then loses off-shell
nilpotency on the antighosts, for which $s\bar c_a\sim b_a$, even
though $s_\psi^2=0$ is true off-shell beforehand.]

If $S$ is linear in the antifields, one may in fact write it as
\be S=S_0 - (s\Phi^A)\Phi_A^*, \ee \noindent by virtue of
\pref{linbrsttrans}.  The gauge fixed version is then \be
S_\psi=S_0-(s_\psi\Phi^A)\frac{\delta\psi}{\delta\Phi^A}=S_0-s_\psi\psi.
\ee \noindent Therefore $s_\psi S_\psi=0$, as the first term is
BRST invariant, and the second term is annihilated by $s_\psi$ by
virtue of $s_\psi^2=0$, which holds off-shell when $S$ is linear
in the antifields as we have just pointed out. The property
$s_\psi S_\psi=0$ is actually quite general and holds even when
$S$ is not linear in the antifields.  It can be proved directly as
follows, \be s_\psi S_\psi=(s_\psi\Phi^A)\frac{\delta^L
S_\psi}{\delta\Phi^A}=-\frac{\delta^R
S}{\delta\Phi_A^*}\frac{\delta^L S_\psi}{\delta\Phi^A}. \ee
\noindent The left-derivative in this expression is a total
derivative, as $S_\psi$ depends on $\Phi_A$ directly and though
the gauge-fixing fermion. Using the chain rule, this yields by
virtue of \pref{replaceds} \be -\frac{\delta^R
S}{\delta\Phi^*_A}\left(\frac{\delta
S}{\delta\Phi^A}+\frac{\delta^2\psi}{\delta\Phi^A\delta\Phi^B}\frac{\delta
S}{\delta\Phi^*_B}\right)=0\label{parityvanish}. \ee \noindent The
second term vanishes because the product of the functional
derivatives of $S$ have a symmetry in ($A$, $B$) opposite to that
of the second functional derivative of $\psi$. The first term
vanishes by the master equation, thus proving the claim.

{}Further information on the gauge-fixed action and the
gauge-fixed cohomology can be found in
\cite{Henneaux:1995ex,Barnich:1999cy,Barnich:2003tr}.

\section{Quantum Master Equation}
\setcounter{equation}{0} \label{quantummaster}

In order to prove gauge independence of the expressions given
above, it is necessary to discuss two important features of the
path integral.
\begin{itemize} \item Assume that a theory of fields $\chi^\alpha$
is given, governed by the action $S[\chi^\alpha]$, with no gauge
invariance (this could be the gauge-fixed action). Expectation
values are, after proper normalization, calculated as \be
<F>=\int{\cal D}\chi
F\exp\left(\frac{i}{\hbar}S[\chi]\right).\label{expvalue} \ee
\noindent The Dyson-Schwinger equations can be directly derived
from the vanishing of the path integral of a total derivative, \be
\int{\cal
D}\chi\frac{\delta}{\delta\chi^\alpha}\left(Fe^{\frac{i}{\hbar}S}\right)=0
\ee \noindent (which is itself a consequence of translation
invariance of the measure). This leads to \be \left<\frac{\delta
F}{\delta\chi^\alpha}+\frac{i}{\hbar}F\frac{\delta
S}{\delta\chi^\alpha}\right>=0, \ee \noindent which is equivalent
to \be \left<F\frac{\delta
S}{\delta\chi^\alpha}\right>=i\hbar\left<\frac{\delta
F}{\delta\chi^\alpha}\right>. \ee \noindent This expression
contains in the l.h.s.\ the expectation values of the classical
equations of motions. In the classical limit $\hbar\to 0$, the
r.h.s. vanishes and the classical equations of motion hold. \item
There is  another aspect that we shall have to take into account.
If $\chi^\alpha$ is changed under a transformation,
$\chi^\alpha\to\chi^\alpha+\epsilon^\alpha$, where
$\epsilon^\alpha$ depends on the fields, the expectation value
\pref{expvalue} is in general not invariant. Furthermore,
invariance of the classical action is not sufficient to guarantee
that the path integral is invariant.  One needs also invariance of
the measure.

We shall be concerned with BRST invariance of the path integral
constructed above.  We have seen that the gauge-fixed action is
BRST invariant.  But the measure might not be.  If it is not, one
may,  in some cases,  restore invariance by taking a different
measure. [The measure is in fact dictated by unitarity and may
indeed not be equal to the trivial measure ${\cal D} \Phi$.]
Invariance is quite crucial in the case of BRST symmetry, since it
is BRST symmetry that guarantees gauge-independence of the
results.  The non-trivial measure terms can be exponentiated in
the action.  Since there is an overall $(1/ \hbar)$ in front of
$S$, the measure terms appear as quantum corrections to $S$.  So,
one replaces the classical action by a ``quantum action" \be
W=S+\hbar M_1+\hbar^2 M_2+...,\label{wexpand} \ee \noindent where
the functionals $M_i$ stem from non-trivial measure factors. The
theorem proved below states that quantum averages are
gauge-independent if the master equation is replaced by the
``quantum master equation" \be \frac{1}{2}(W,W)=i\hbar\Delta
W,\label{qme} \ee \noindent where $\Delta$ is defined in
\pref{defdelta}. Note that if $\Delta S=0$, the Jacobian is unity
for the BRST transformation and $W$ might then be taken equal to
$S$. The quantum master equation reduces to the classical master
equation considered above, which is solved by $S$.  While there is
always a solution to the classical master equation, the solution
to the quantum master equation might get obstructed.  We shall
investigate this question below.  For the moment, we assume that
there is no obstruction.
\end{itemize}

We can now state the correct, general rules, for computing
expectation values of observables (including $1$):  these are the
quantum averages, weighted by $\exp (\frac{i}{\hbar}W)$, of the BRST
observables corrected by the addition of appropriate $\hbar$ (and
possibly also higher) order terms. Namely, consider a classical
observable $A_0$. Construct its (in fact, one of its) BRST-invariant
extension $A = A_0 + \hbox{ghost terms}$, so that $(S,A) = 0$. The
BRST cocycle $A$ has to be augmented as \be A\to\alpha=A_0+\hbar
B_1+\hbar^2 B_2+... \ee \noindent where the terms of order $\hbar$
and higher must be such that $\sigma\alpha=0$, where $\sigma$ was
defined in the exercises as \be
\sigma\alpha\equiv(W,\alpha)-i\hbar\Delta\alpha, \ee \noindent with
$W$ the solution of the quantum master equation. (Note that these
$B$-terms come over and above the ghost terms needed classically to
fulfill $(S,A) = 0$.) The operator $\sigma$ is the quantum
generalization of $s$. The $\psi$-independent expectation value
$<A_0>$ of the observable $A_0[\phi^i]$ is computed from $\alpha$ as
\be <A_0>=\int{\cal
D}\Phi^A\alpha\left(\Phi^A,\Phi_A^*=\frac{\delta\psi}{\delta\Phi^A}\right)
\exp\left(\frac{i}{\hbar}W\left[\Phi^A,\Phi^*_A=\frac{\delta\psi}
{\delta\Phi^A}\right]\right).\label{qexpvalue} \ee \noindent The
claim is that this expectation value does not depend on the choice
of the gauge-fixing fermion \bea
\psi'=\psi+\delta\psi\equiv\psi+\mu\label{gfvariation}\\
<A_0>_{\psi'}=<A_0>_\psi, \eea \noindent where $\mu$ is an
arbitrary modification of $\psi$.

To prove the claim, we denote the argument of the integral by $V$
for convenience in the following.  It has been already shown in
the exercises that \be \Delta V=0\Leftrightarrow\sigma\alpha=0.
\ee Now, perform the variation of the gauge fixing functional
\pref{gfvariation}. The variation of the quantum average
\pref{qexpvalue} is equal to \be \int{\cal
D}\Phi\frac{\delta^L\mu}{\delta\Phi^A}\frac{\delta^LV}{\delta\Phi_A^*}.\label{gi}
\ee \noindent To evaluate this expression, note that \be
\frac{\delta^L}{\delta\Phi^A}\left(\mu\frac{\delta^L
V}{\delta\Phi_A^*}\right)=\frac{\delta^L\mu}{\delta\Phi^A}
\frac{\delta^LV}{\delta\Phi_A^*}+(-1)^{\epsilon_A}\frac{\delta^L\delta^L
V}{\delta\Phi^A\delta\Phi_A^*}. \ee \noindent The derivatives are
total ones. Denoting partial derivatives by a prime ', the last
term can be rewritten as \be \frac{\delta^L\delta^L
V}{\delta\Phi^A\delta\Phi_A^*}=\frac{\delta^{'L}\delta^L
V}{\delta'\Phi^A\delta\Phi_A^*}+\frac{\delta^2\psi}
{\delta\Phi^A\delta\Phi^B}\frac{\delta^L\delta^L
V}{\delta\Phi_B^*\delta\Phi_A^*}=\Delta V=0. \ee \noindent As in
the case of equation \pref{parityvanish}, the second term vanishes
by parity arguments. Thus the integral \pref{gi} can be rewritten
as a total derivative in field space, which vanishes in view of
translation invariance of the standard measure. Therefore the path
integral does not get modified if one changes the gauge-fixing
fermion, as claimed.

Given $A_0$, its BRST extension is determined up to a BRST exact
term $s B$, see \pref{genobj}.  This ambiguity can be extended to
higher orders in $\hbar$ as \be \alpha\to\alpha+\sigma\beta. \ee
\noindent It has been shown in the exercises that \be
(\sigma\beta)\exp\left(\frac{i}{\hbar}W\right)
\sim\Delta\left(\beta\exp\left(\frac{i}{\hbar}W\right)\right). \ee
\noindent As the r.h.s\ is a total derivative by its definition,
\pref{defdelta}, the path-integral over the l.h.s.\ vanishes \be
\int{\cal D}\Phi\sigma\beta\exp\left(\frac{i}{\hbar}W\right)=0.
\ee \noindent Therefore adding any element in the image of
$\sigma$ does not alter the quantum averages.  The path integral
associates a unique answer to a given cohomological class of
$\sigma$, i.e., does not depend on the choice of representative.

Note that the ambiguity in $\alpha$, given $A_0$, is more than
just adding a $\sigma$-trivial term to $\alpha$. At each order in
$\hbar$ one may add a non trivial new observable since this does
not modify the classical limit.  This addition is relevant, in the
sense that it changes the expectation value by terms of order
$\hbar$ or higher. This is an unavoidable quantum ambiguity.  A
similar ambiguity exists for the quantum measure (i.e., the $M_1$,
$M_2$ etc.\ terms in $W$).  These terms do not spoil BRST invariance
and must be determined by other criteria, e.g.,  by comparison
with the Hamiltonian formalism.

\section{Anomalies}
\setcounter{equation}{0} \label{anomalies} We close this brief
survey by analyzing the possible obstructions to the existence of
a solution $W$ to the quantum master equation.  This leads to the
important concept of anomalies. The fact that anomalies in the
Batalin-Vilkovisky formalism appear as an incurable violation of
the BRST invariance of the measure was first investigated in
\cite{Troost:1989cu}.

Analyzing the obstructions to the existence of a solution to the
quantum master equation can be done most easily from a direct
$\hbar$ expansion. To order $\hbar^0$ one gets from the quantum
master equation \be \frac{1}{2}(S,S)=0, \ee \noindent which is the
classical master equation.  This equation is certainly fulfilled,
since there is no obstruction to the existence of $S$.

To the next order $\hbar$, the quantum master equation yields \be
sM_1=(S,M_1)=i\Delta S.\label{dsexact} \ee \noindent Given the
$\hbar^0$-term $S$, this equation has a solution for $M_1$ if
$s\Delta S=(S,\Delta S)=0$. This condition is necessary but in
general not sufficient (see below). To prove that the condition
$(S,\Delta S)=0$ holds, we note that \be
\Delta(\alpha,\beta)=(\Delta\alpha,\beta)-(-1)^{\epsilon_\alpha}(\alpha,\Delta\beta),
\label{keycentral}\ee \noindent as was proven in the exercises.
This property uses $\Delta^2=0$ and the generalized Leibniz rule
\be \Delta(\alpha\beta)=(\Delta\alpha)\beta+
(-1)^{\epsilon_\alpha}\alpha\Delta\beta+(-1)^{\epsilon_\alpha}(\alpha,\beta).
\ee \noindent For $\alpha=\beta=S$, the l.h.s\ of
(\ref{keycentral}) vanishes by $(S,S)=0$. In view of the gradings
of $S$, the r.h.s\ yields $2(\Delta S,S)$. Thus, $(\Delta S,S) =
0$, that is,  $\Delta S$ is closed, as requested.

This does not imply that $\Delta S$ is exact, however, unless the
cohomological group  $H^1(s)$ vanishes at ghost number one (recall
that $\Delta S$ has ghost number one). But \pref{dsexact} requires
$\Delta S$ to be exact. If $\Delta S$ is not exact, there is no
$M_1$ and therefore, no way to define a BRST invariant measure
such that the quantum averages do not depend on the gauge fixing
fermion.  This presumably signals a serious pathology of the
theory. If $\Delta S$ is exact, $M_1$ exists and one can
investigate the problem of existence of the next term $M_2$. One
easily verifies that it is again $H^1(s)$ that measures the
potential obstructions to the existence of this next term $M_2$,
as well as the existence of the subsequent terms $M_3$ etc.

In particular, if one can show that $H^1(s)$ vanishes, one is
guaranteed that a solution of the quantum master equation exists.
If $H^1(s) \not= 0$, further work is required since one must check
that one does not hit an obstruction.  Note that $W$ should be a
local functional (with possibly infinite coupling constants), so
that the relevant space in which to compute the cohomology is that
of local functionals.

The computation of the local cohomology of the BRST operator for
Yang-Mills gauge theory has been carried out in
\cite{Barnich:1994ve,Barnich:1994db,Barnich:1994mt}, following
earlier work without antifields
\cite{Brandt:1989gv,Brandt:1989rd,Brandt:1989gy,Dubois-Violette:1985hc,Dubois-Violette:1985jb,Dubois-Violette:1992yy}.
See \cite{Barnich:2000zw} for a review.

\section*{Acknowledgements}
We are grateful to the organizers of the $10^{th}$ Saalburg Summer
School on Modern Theoretical Methods for their kind invitation and
hospitality.  We are also grateful to Jim Stasheff for a careful
reading of the manuscript.  The work of MH is supported in part by
the ``Interuniversity Attraction Poles Programme -- Belgian Science
Policy '', by IISN-Belgium (convention 4.4505.86) and by the
European Commission FP6 programme MRTN-CT-2004-005104, in which he
is associated to the V.U.Brussel (Belgium).
\\
\appendix
\setcounter{equation}{0} \no
\section{Appendix I: De Witt's notation.}

\no We review the De Witt's condensed notation. This notation makes
it possible to write gauge transformations in a more compact form:
\be \delta_{\varepsilon}\varphi^i=R^i_{\alpha}\varepsilon^{\alpha}
\leftrightarrow \delta_{\varepsilon}\varphi^i(x)=\int d^ny
\;R^i_{\alpha}(y,x)\varepsilon^{\alpha}(y). \ee \no For example, the
transformation of the Yang-Mills gauge field \be
\delta_{\varepsilon}A_{\mu}^a=D_{\mu}\varepsilon^a
=\partial_{\mu}\varepsilon^a+f^a_{\;\;cb}A_{\mu}^b\varepsilon^c \ee
\no can be written as \be \delta_{\varepsilon}A_{\mu}^a=R_{\mu b}^a
\varepsilon^b,\;\;\;\;R_{\mu
b}^a(x,y)=\partial_{\mu}\delta(x-y)\delta_b^a+f^a_{\;\;bc}A_{\mu}^b\delta(x-y).
\ee \no Noether's identities have the simple form: \be \frac{\delta
S_0}{\delta \varphi^i}R^i_{\alpha}=0. \ee
\\

\no \section{Appendix II: Properties of anti-brackets.}
\setcounter{equation}{0}

\no (i) $(F,G)=-(-1)^{(\varepsilon_F+1)(\varepsilon_G+1)}(G,F)$,
where $\varepsilon_F=0\;(1)$ for $F$ bosonic (fermionic);
the anti-bracket is symmetric if both $F$ and $G$ are bosonic,
and antisymmetric otherwise. \\

\no (ii) Jacobi identitiy:\\
\indent $(-1)^{(\varepsilon_F+1)(\varepsilon_H+1)}(F,(G,H))+\mbox{cyclic permutations}=0$ \\

\no (iii) $(FG,H)=F(G,H)+(F,H)G \;(-1)^{\varepsilon_G(\varepsilon_H+1)}$; \\
\indent $\;\;(F,GH)=(F,G)H+G(F,H) \;(-1)^{\varepsilon_G(\varepsilon_F+1)}$ \\

\no (iv) $\gh\left((F,G)\right)=\gh F+\gh G+1$ \\

\addcontentsline{toc}{section}{References}
\bibliographystyle{utphys}
\bibliography{notes}

\end{document}